\documentclass[pra,twocolumn,superscriptaddress,showpacs,amsmath,amssymb]{revtex4-1}
\usepackage{epsfig}
\usepackage{subfigure}
\usepackage{amssymb,amsmath}
\usepackage{graphicx}
\usepackage{epstopdf}
\usepackage{epsfig}
\usepackage{xcolor}
\usepackage{color}
\usepackage{amsfonts}
\usepackage{amscd}
\usepackage{hyperref}
\usepackage{bbm}
\usepackage{mathrsfs}
\DeclareMathSizes{10}{10}{7}{5}

\begin{document}
\title{Quantum Langevin approach for non-Markovian quantum dynamics \\of the spin-boson model}
\date{\today}
\author{Zheng-Yang Zhou}
\thanks{Z.Y.Z and M.C. contributed equally to this work.}
\affiliation{Department of Physics, Fudan University, Shanghai 200433, China}
\affiliation{Beijing Computational Science Research Center, Beijing
100094, China}
\author{Mi Chen}
\thanks{Z.Y.Z and M.C. contributed equally to this work.}
\affiliation{Department of Physics, Fudan University, Shanghai 200433, China}
\affiliation{Beijing Computational Science Research Center, Beijing
100094, China}
\author{Ting Yu}
\affiliation{Beijing Computational Science Research Center, Beijing
100094, China}
\affiliation{Center for Controlled Quantum Systems and Department of Physics and Engineering Physics,
Stevens Institute of Technology, Hoboken, New Jersey 07030, USA}
\author{J. Q. You}
\altaffiliation[jqyou@csrc.ac.cn]{}
\affiliation{Beijing Computational Science Research Center, Beijing
100094, China}

\begin{abstract}

One long-standing difficult problem in quantum dissipative dynamics is to solve the spin-boson model in a non-Markovian regime where a tractable systematic master equation does not exist. The spin-boson model is particularly important due to its crucial applications in quantum noise control and manipulation as well as its central role in developing quantum theories of open systems.  Here we solve this important model by developing a non-Markovian quantum Langevin approach. By projecting the quantum Langevin equation onto the coherent states of the bath, we can derivie a set of non-Markovian quantum Bloch equations containing no explicit noise variables. This special feature offers a tremendous advantage over the existing stochastic Schr\"{o}dinger equations in numerical simulations. The physical significance and generality of our approach are briefly discussed.
 \end{abstract}
\maketitle

\section{Introduction}

Quantum Langevin equation (QLE) provides a direct depiction of the temporal behaviors of physical observables under the influence of a bath of quantum particles~\cite{qle1,new1,new2,qle1.5,qle3,qle4}.
As such, QLE has many important applications in quantum optics \cite{new1}, the input-output theory~\cite{new3}, and the quantum dynamics of dissipative
atoms~\cite{qle6,qle6.5,qle6.75,qle7}.  For deriving a generic Langevin equation,  however, Markovian approximation was usually employed
to arrive at a tractable equation of motion.  QLE beyond Markovian approximation can be also formulated to study the intriguing non-Markovian
dynamics of damped quantum systems and the Brownian motion systems \cite{,qle7.5,qle8,qle9,qle10,qle11,qle12}. A method\cite{Gri1,Gri2,Gri3} based on the Mori expansion\cite{Mori1,Mori2} can solve these kind of problems conveniently. The main idea of this method is to expand the time dependent operator with a set of time independent basis operators. This set of basis and the corresponding coefficients are govern by two recurrence relations.

In the last decade, the so-called non-Markovian quantum state diffusion (QSD) equation~\cite{Diosi,Strunz1,Yu1} has been formulated nonperturbatively,
so it can  apply to the cases with strong couplings between systems and environments (see, e.g., \cite{Wiseman,Strunz2,Jing1}). The non-Markovian
QSD has provided a powerful tool in numerically simulating many interesting physical models \cite{N-level,N-qubit}. In particular,
high-order numerical methods for the non-Markovian QSD~\cite{Strunz3,ZCL,dwl} have been developed very recently, making some previously intractable problems becoming numerically tractable. In fact, the QSD equation is a stochastic Schr{\"o}dinger equation and it is solved by invoking the noise realizations.
For the important case of spin-boson model \cite{sb1}, however, matters
are not as simple as the form of this model due to the fact that the spin-boson model does not admit an analytical treatment and an efficient numerical simulation
is prohibited  without including the higher-order perturbations \cite{Strunz3,ZCL,dwl}.

In this paper, we develop a stochastic quantum Langevin approach to solving non-Markovian quantum dynamics of the spin-boson model.
This model which is the multi-mode case of the quantum Rabi model\cite{rabi1,rabi2,rabi3} involves non-conserving processes due to the counter-rotating terms. Consequently, it poses a long-standing difficult problem
in studying non-Markovian quantum dynamics~\cite{sb1}. By projecting the non-Markovian QLE onto the coherent states of the bath, we convert the operator QLE into a c-number stochastic QLE, which is formally analogous to the non-Markovian QSD equation. Therefore, the useful techniques
developed for the QSD can apply to the c-number stochastic QLE as well.  Remarkably, we find that the stochastic QLE can be further reduced to a set of simple non-Markovian quantum Bloch equations without involving any noise variables. This provides a much more efficient method
to solve the non-Markovian  quantum dynamics of the spin-boson model. As shown below, the method developed here is quite general, so it may offer significant numerical advantages for simulating open quantum systems coupled to bosonic environments when higher-order perturbation is unavoidable.

The paper is organized as follows. In Sec.~II, we obtain a stochastic QLE by projecting the non-Markovian QLE onto the coherent states of the bosonic bath. Then, in Sec.~III, we convert the stochastic QLE into a c-number stochastic QLE, which is formally analogous to the non-Markovian QSD equation.
In Sec.~IV, we further reduce the c-number stochastic QLE to a set of simple non-Markovian quantum Bloch equations.  The Extensions to the cases of complex correlation function and finite temperature are discussed in Sec.~V and VI, respectively. Finally, Sec.~VII gives the conclusion of our work.

\section{Stochastic QLE}

The spin-boson model is described by $H_{\rm tot}=H_{\rm 0}+H_{\rm int}$, with (setting $\hbar=1$)
\begin{eqnarray}
H_0 &=& \frac{\omega}{2}\sigma_z+\sum_k\omega_ka^{\dag}_ka_k,\nonumber\\
H_{\rm int} &=& \sigma_x\sum_k\left(g_{k}^*a_k^{\dag}+g_{k}a_k\right).
\label{H}
\end{eqnarray}
Here $H_0$ is the Hamiltonian of the uncoupled spin and multi-mode bosonic bath, $H_{\rm int}$ models the interaction between the spin and the bosonic bath, $\sigma_x$, $\sigma_y$ and $\sigma_z$ are Pauli operators, and $a_{ k}^\dagger$ ($a_{ k}$) is the $k$th-mode bosonic creation (annihilation) operator of the bath. We assume that the state of the total system is initially factorized as
$|\Psi_0\rangle=|\psi\rangle\otimes|0\rangle$, where the bosonic bath is in the vacuum state $|0\rangle$ (i.e., at zero temperature).

The interaction Hamiltonian $H_{\rm int}$ can be rewritten as the sum of rotating and counter-rotating terms,
%\begin{equation}
\[
H_{\rm int}=\sum_k (g_{k}^*a_k^{\dag}\sigma_-+g_{k}a_k\sigma_++g_{k}^*a_k^{\dag}\sigma_++g_{k}a_k\sigma_- ),
\]
%\end{equation}
with $\sigma_x=\sigma_+ +\sigma_-$. The counter-rotating terms  $g_{k}^*a_k^{\dag}\sigma_+$ and $g_{k}a_k\sigma_-$  break the conservation of excitation number,
giving rise to high-order noise appearing in the stochastic equation of quantum dynamics \cite{Diosi,Strunz1,Yu1}.

Starting from the Heisenberg equations of the Pauli operators and the bosonic operators of the bath, one obtains

%The beginning of the red text
\begin{eqnarray}
\frac{d}{dt}\sigma_x(t)&=&-\omega\sigma_y(t),\nonumber\\
\frac{d}{dt}\sigma_y(t)&=&\omega\sigma_x(t)-2\sum_k\sigma_z(t)\left[g_ka_k(t)+g_k^*a_k^{\dag}(t)\right],\nonumber\\
\frac{d}{dt}\sigma_z(t)&=&2\sum_k\sigma_y(t)\left[g_ka_k(t)+g_k^*a_k^{\dag}(t)\right],\nonumber\\
\frac{d}{dt}a_k(t)&=&-i\omega_ka_{k}(t)-ig_k^*\sigma_x(t).\nonumber\\
\label{HE1}
\end{eqnarray}
where
%\begin{equation}
$$
\sigma_j(t)=e^{iH_{\rm tot}t}\sigma_je^{-iH_{\rm tot} t},~~j=x,y,z,
$$
%\end{equation}

are Pauli operators in the Heisenberg picture, and
\begin{eqnarray}
a_k(t)&=&e^{iH_{\rm tot}t}a_ke^{-iH_{\rm tot} t},\nonumber
\end{eqnarray}
are the field operators in the Heisenberg picture.
The field operators $a_k(t)$ and $a_k^{\dag}(t)$ in Eq.~$(\ref{HE1})$ can be formally solved

\begin{eqnarray}
a_k(t)&=&e^{-i\omega_kt}a_k-ig_k^*\int_0^tdse^{-i\omega_k(t-s)}\sigma_x(s).\nonumber\\
\label{HE2}
\end{eqnarray}
Substituting the field operators in Eq.~$(\ref{HE1})$ with the formal solution in Eq.~$(\ref{HE2})$ we can achieve the QLE,

\begin{eqnarray}
\frac{d}{dt}\sigma_x(t) &=& -\omega\sigma_y(t),\nonumber\\
\frac{d}{dt}\sigma_y(t) &=& \omega\sigma_x(t)-2\sigma_z(t)\left[\xi(t)+\xi^\dag(t)\right]+w_z(t),\nonumber\\
\frac{d}{dt}\sigma_z(t) &=& 2\sigma_y(t)\left[\xi(t)+\xi^\dag(t)\right]-w_y(t),
\label{QLE1}
\end{eqnarray}
with $\alpha(t,s)\equiv\sum_k|g_k|^2e^{-i\omega(t-s)}$ is the correlation function of the bath,
$\xi(t)\equiv \sum_{k}g_ke^{-i\omega_kt}a_k$ defines a noise operator, and
%\begin{equation}
\[
w_j(t)\equiv2i\sigma_j(t)\int_0^tds[\alpha(t,s)-\alpha^*(t,s)]\sigma_x(s),
\]
%\end{equation}
with $j=y,z$.
In this QLE, both $\xi(t)$ and $\xi^\dag(t)$ act as ``random noises" acting on the spin.

Here we first consider the case of real correlation function $\alpha(t,s)=\alpha^*(t,s)$, so that $w_y(t)=w_z(t)=0$ in Eq.~(\ref{QLE1}). Note that the typical Ornstein-Uhlenbeck correlation function $\alpha(t,s)=\frac{\Gamma\gamma}{2}e^{-\gamma|t-s|}$ is indeed a real function.
We define Bargmann coherent states for the bosonic bath,
\begin{equation}
|z\rangle\equiv \bigotimes_{k}| z_{k}\rangle=e^{\sum_{k}z_{k}a_{k}^\dagger}|0\rangle,
\end{equation}
which satisfy
\begin{equation}
a_k|z\rangle=z_k|z\rangle,~~~a_k^{\dag}|z\rangle=\frac{\partial}{\partial z_k}|z\rangle.
\end{equation}
When projected onto the Bargmann coherent states, the QLE in Eq.~(\ref{QLE1}) is then converted to
\begin{eqnarray}
\frac{\partial}{\partial t}\sigma_x(t;z) &=& -\omega\sigma_y(t;z),\nonumber\\
\frac{\partial}{\partial t}\sigma_y(t;z) &=& \omega\sigma_x(t;z)\nonumber\\
                                         &&-2\left[z_t+\int_0^tds\alpha(t,s)\frac{\delta}{\delta z_s}\right]\sigma_z(t;z),\nonumber\\
\frac{\partial}{\partial t}\sigma_z(t;z) &=& 2\left[z_t+\int_0^tds\alpha(t,s)\frac{\delta}{\delta z_s}\right]\sigma_y(t;z).
\label{sQLE2}
\end{eqnarray}
This is a stochastic QLE with the noise $z_t=\sum_{k}g_ke^{-i\omega_kt}z_k$.
In Eq.~(\ref{sQLE2}), $\sigma_j(t;z)\equiv{\sigma}_j(t)|z\rangle\langle{z}|$, with $j=x,y,z$, and the functional chain rule,
%\begin{equation}
\[
\frac{\partial}{\partial z_k}=\int ds\frac{\partial z_s}{\partial z_k}\frac{\delta}{\delta z_s},
\]
%\end{equation}
is used. Note that
\begin{equation}
\sigma_j(t)=\prod_{k}\int \frac{d^2z_k}{\pi}e^{-| z_{k}|^2}\sigma_j(t;z)\equiv\mathcal{M}\left\{\sigma_j(t;z)\right\}.
\label{completeness}
\end{equation}
When statistically averaging Eq.~(\ref{sQLE2}) over all noise variables via Eq.~(\ref{completeness}), one can recover Eq.~(\ref{sQLE2}) back to the QLE in Eq.~(\ref{QLE1}).

\section{C-number stochastic QLE}

To convert the stochastic equation of operators [i.e., Eq.~(\ref{sQLE2})] into a c-number equation, we introduce the expectation value of an operator $\sigma$ as $\langle{\sigma}\rangle\equiv\langle\Psi_0|\sigma|\Psi_0\rangle$. Then, we have
\begin{eqnarray}
\frac{\partial}{\partial t}\langle\sigma_x(t;z)\rangle &=& -\omega\langle\sigma_y(t;z)\rangle,\nonumber\\
\frac{\partial}{\partial t}\langle\sigma_y(t;z)\rangle &=& \omega\langle\sigma_x(t;z)\rangle\nonumber\\
                                                       &&-2\left[z_t+\int_0^tds\alpha(t,s)\frac{\delta}{\delta z_s}\right]\langle\sigma_z(t;z)\rangle,\nonumber\\
\frac{\partial}{\partial t}\langle\sigma_z(t;z)\rangle &=& 2\left[z_t+\int_0^tds\alpha(t,s)\frac{\delta}{\delta z_s}\right]\langle\sigma_y(t;z)\rangle.
\label{SCNQLE1}
\end{eqnarray}
Define $\mathcal{A}(t,z)\equiv[\langle\sigma_x(t;z)\rangle,\langle\sigma_y(t;z)\rangle,\langle\sigma_z(t;z)\rangle]^T$, where $T$ denotes the transpose of a matrix. Equation (\ref{SCNQLE1}) can be written in a matrix form as
\begin{eqnarray}
\frac{\partial}{\partial t}\mathcal{A}(t,z)
               &=& -i\mathcal{H}\mathcal{A}(t,z)+\mathcal{L}z_t\mathcal{A}(t,z)\nonumber\\
               &&+\mathcal{L}\int_0^tds\alpha(t,s)\frac{\delta}{\delta z_s}\mathcal{A}(t,z),
\label{SCNQLE2}
\end{eqnarray}
with
\begin{equation}
\mathcal{H}=\left(\begin{array}{ccc}
               0&-i\omega&0\\
               i\omega&0&0\\
               0&0&0\\
               \end{array}\right),~~
\mathcal{L}=\left(\begin{split}%{\nonumber}
               0&&0&&0\\
               0&&0&&-2\\
               0&&2&&0\\
              \end{split}\right).
\end{equation}
Formally, this c-number stochastic QLE is analogous to the non-Markovian QSD equation governing non-Markovian quantum trajectories~\cite{Diosi}. The difference here is that the QSD equation is a stochastic differential equation for quantum states of the system (i.e., a stochastic Schr{\"o}dinger equation), while the c-number stochastic QLE in Eq.~(\ref{SCNQLE2}) corresponds to a stochastic differential equation of physical variables.

Here we introduce  $\mathcal{O}(t,s,z)$ operator by
\begin{equation}
\frac{\delta}{\delta z_s}\mathcal{A}(t,z)=\mathcal{O}(t,s,z)\mathcal{A}(t,z).
\end{equation}
Note that although we use the notation  $\mathcal{O}(t,s,z)$ which is similar to the $O$ operator used
in QSD approach, their meanings are different.
Here $\mathcal{O}(t,s,z)$ is defined for an arbitrary operator, rather than for a quantum state.  Now we can write the c-number
stochastic QLE in a time-local form,
\begin{equation}
\frac{\partial}{\partial t}\mathcal{A}(t,z)
               =\left[-i\mathcal{H}+\mathcal{L}z_t+\mathcal{L}\bar{\mathcal{O}}(t,z)\right]\mathcal{A}(t,z),
\label{SCNQLE3}
\end{equation}
where
%\begin{equation}
\[
\bar{\mathcal{O}}(t,z)=\int_0^tds\alpha(t,s)\mathcal{O}(t,s,z).
\]
%\end{equation}
Also, using Eq.~(\ref{SCNQLE3}) and the relation
%\begin{equation}
\[
\frac{\delta}{\delta z_s}\frac{\partial}{\partial t}\mathcal{A}(t,z)=
\frac{\partial}{\partial t}\frac{\delta}{\delta z_s}\mathcal{A}(t,z),
\]
%\end{equation}
we obtain the equation for $\mathcal{O}(t,s,z)$ operator,
 \begin{eqnarray}
\frac{\partial}{\partial t}\mathcal{O}(t,s,z) &=& [-i\mathcal{H}+\mathcal{L}z_t+\mathcal{L}\bar{\mathcal{O}}(t,z),\mathcal{O}(t,s,z)]\nonumber\\
                     &&+\mathcal{L}\frac{\delta\bar{\mathcal{O}}(t,z)}{\delta z_s}.
\label{eforO}
\end{eqnarray}
As in Ref.~\onlinecite{Wiseman}, the initial condition of the $\mathcal{O}(t,s,z)$ operator can be derived as
$\mathcal{O}(t,t,z)=\mathcal{L}$.

\section{Non-Markovian quantum Bloch equation}

To obtain the desired quantity
\begin{eqnarray}
\mathcal{A}(t)&=&\mathcal{M}\{\mathcal{A}(t,z)\} \nonumber \\
&\equiv&[\langle\sigma_x(t)\rangle,\langle\sigma_y(t)\rangle,\langle\sigma_z(t)\rangle]^T, \label{aoaz}
\end{eqnarray}
where
$\langle\sigma_j(t)\rangle=\langle\mathcal{M}\{\sigma_j(t;z)\}\rangle$, one can numerically solve Eq.~(\ref{SCNQLE3}) for each realization of the noise $z_t$ and then implement the statistical average, as in the case of numerically solving QSD equation. However, when higher-order perturbation is involved in the QSD,  one must pay the price of long computation time in order to achieve accurate results. Below we show that with our QLE approach, this simulation process can be significantly sped up.

By directly implementing statistical average on Eq.~(\ref{SCNQLE3}), we have
\begin{eqnarray}
\frac{\partial}{\partial t}\mathcal{A}(t)
&=& -i\mathcal{H}\mathcal{A}(t)+\mathcal{L}\mathcal{M}\left\{z_t\mathcal{A}(t,z)\right\} \nonumber\\
&&\!+\mathcal{L}\mathcal{M}\left\{\bar{\mathcal{O}}(t,z)\mathcal{A}(t,z)\right\}.
\label{CNQLE1}
\end{eqnarray}
In Eq.~(\ref{CNQLE1}), $\mathcal{M}\left\{z_t\mathcal{A}(t,z)\right\}$ can be written as
$$\mathcal{M}\{z_t\mathcal{A}(t,z)\}=\prod_{k}\int \frac{d^2z_k}{\pi}e^{-| z_{k}|^2}z_t\langle\Psi_0|\mathcal{B}(t)|z\rangle\langle z|\Psi_0\rangle,$$
where $\mathcal{B}(t)\equiv(\sigma_x(t),\sigma_y(t),\sigma_z(t))^T$.
Because $z_t|z\rangle=\xi(t)|z\rangle$, we have
\begin{eqnarray}
\mathcal{M}\{z_t\mathcal{A}(t,z)\} &=& \langle\Psi_0|\mathcal{B}(t)\xi(t) \nonumber\\
&&~~\times\prod_{k}\int \frac{d^2z_k}{\pi}e^{-| z_{k}|^2}|z\rangle\langle z|\Psi_0\rangle \nonumber\\
                                   &=& \langle\Psi_0|\mathcal{B}(t)\xi(t)|\Psi_0\rangle \nonumber\\
                                   &=& 0,
\label{zterm}
\end{eqnarray}
where we have used the relation $\xi(t)|\Psi_0\rangle=\sum_{k}g_ke^{-i\omega_kt}a_k|\psi\rangle\otimes|0\rangle=0$.

It is known that the $\mathcal{O}(t,s,z)$ operator can be expanded as~\cite{Yu1}
\begin{eqnarray}
\mathcal{O}(t,s,z) &=& \mathcal{O}_0(t,s)
                   +\sum_{n(\geq1)}\int_0^t\mathcal{O}_{n}(t,s,v_1,\ldots,v_n) \nonumber\\
                   & &\!~~~~~~~~~~~~~~~~~~~\times z_{v_1}\ldots z_{v_n}dv_1\ldots dv_n.~~~~
\label{Oexp}
\end{eqnarray}
Because
\begin{eqnarray}
&&\mathcal{M}\left\{\bar{\mathcal{O}}_{n}(t,v_1,\ldots,v_n)z_{v_1}\ldots z_{v_n}\langle\Psi_0|\mathcal{B}(t)|z\rangle\langle z|\Psi_0\rangle\right\}  \nonumber\\
       &&=\mathcal{M}\left\{\bar{\mathcal{O}}_{n}(t,v_1,\ldots,v_n)\langle\Psi_0|\mathcal{B}(t)\xi(v_1)\ldots \xi(v_n)|z\rangle\langle z|\Psi_0\rangle\right\} \nonumber\\
       &&=\bar{\mathcal{O}}_{n}(t,v_1,\ldots,v_n)\langle\Psi_0|\mathcal{B}(t)\xi(v_1)\ldots \xi(v_n)\mathcal{M}\left\{|z\rangle\langle z|\right\}\Psi_0\rangle \nonumber\\
       &&=\bar{\mathcal{O}}_{n}(t,v_1,\ldots,v_n)\langle\Psi_0|\mathcal{B}(t)\xi(v_1)\ldots \xi(v_n)|\Psi_0\rangle \nonumber\\
       &&=0,
\end{eqnarray}
where
%\begin{equation}
\[
\bar{\mathcal{O}}_{n}(t,v_1,\ldots,v_n)=\int_0^tds\alpha(t,s)\mathcal{O}_n(t,s,v_1,\ldots,v_n),
\]
%\end{equation}
Eq.~(\ref{CNQLE1}) is finally reduced to our central result
\begin{equation}
\frac{\partial}{\partial t}\mathcal{A}(t)=-i\mathcal{H}\mathcal{A}(t)+\mathcal{L}\bar{\mathcal{O}}_0(t)\mathcal{A}(t).
\label{exact}
\end{equation}
Here we call it a non-Markovian quantum Bloch equation, in which no noise variables are involved. Now note that only the noiseless term of the functional expansion in Eq.~(\ref{Oexp})
is important in solving the non-Markovian quantum dynamics of the system. Because no noise variables are involved, Eq.~(\ref{exact}) can be numerically solved very efficiently.

Figure~1 shows the time evolution of $\langle\sigma_z\rangle$ for a bath with the Ornstein-Uhlenbeck correlation function. It can be seen from
Fig. 1(a) that when increasing $\gamma$ , the environmental memory
time $1/{\gamma}$ decreases and $\langle\sigma_z\rangle$ exhibits a clear transition from
an oscillation to an exponential decay. Physically, this is to
some extent connected to the overdamped oscillator, where
increasing the friction on the velocity of the oscillator has the
effect of turning an oscillation into an exponential decay.
%$\alpha(t,s)=\frac{\Gamma\gamma}{2}e^{-\gamma|t-s|}$.

The results are obtained by solving the non-Markovian quantum Bloch equation (\ref{exact}), with $\bar{\mathcal{O}}_0(t)$ determined by (see Appendix A)
\begin{eqnarray}
\frac{\partial}{\partial t}{\bar{\mathcal{O}}}_{0}(t) &=& -i[\mathcal{H},\bar{\mathcal{O}}_{0}(t)]
+[\mathcal{L}\bar{\mathcal{O}}_{0}(t),\bar{\mathcal{O}}_{0}(t)]
-\gamma\bar{\mathcal{O}}_{0}(t) \nonumber\\
&& +\frac{\Gamma\gamma}{2}\mathcal{L}+\mathcal{L}\mathcal{Q}_{1},\nonumber\\
\frac{\partial}{\partial t}\mathcal{Q}_{n}
&=& -i[\mathcal{H},\mathcal{Q}_{n}]
+\sum_{k=0}^n[\mathcal{L}\mathcal{Q}_{k},\mathcal{Q}_{n-k}]-(n+1)\gamma \mathcal{Q}_{n} \nonumber\\
&& +\frac{\Gamma\gamma}{2}[\mathcal{L},\mathcal{Q}_{n-1}]+(n+1)\mathcal{L}\mathcal{Q}_{n+1},
\label{eforQn}
\end{eqnarray}
with initial condition $\mathcal{Q}_{0}=\bar{\mathcal{O}}_{0}(t)$. These hierarchical equations do not contain any explicit noise variables. In numerical calculations, one can truncate Eq.~(\ref{eforQn}) at a given hierarchical order $\mathcal{N}$ by choosing $\mathcal{Q}_{\mathcal{N}+1}=0$. The results in Fig.~1(a) are very similar to those in Ref.~\cite{ZCL} obtained using the QSD method, showing apparent non-Markovian behaviors at small values of $\gamma$.  In \cite{ZCL}, the simulations for the curve with $\gamma=0.2$ took about 36 days to execute on an Intel core-i7 CPU core, but only a few seconds here by solving the non-Markovian quantum Bloch equation (\ref{exact}) via the noiseless hierarchical equations in Eq.~(\ref{eforQn}). This is because of the numerical efficiency of our method without invoking any noise realizations. While $\langle\sigma_z\rangle$ at $\mathcal{N}=0$ and 3 deviate from those at $\mathcal{N}=10$ and 100, $\langle\sigma_z\rangle$ at $\mathcal{N}=10$ and 100 look nearly identical [see Fig.~1(b)], revealing fast convergence of our results with the hierarchical order $\mathcal{N}$. In contrast, the results of $\langle\sigma_z\rangle$ obtained using the QSD method have considerable differences between the $\mathcal{N}=10$ and 100 orders of the hierarchical equation (see Fig.~2 in \cite{ZCL}), indicating much slower convergence with $\mathcal{N}$ there.

\begin{figure}
\includegraphics[width=3.4in]{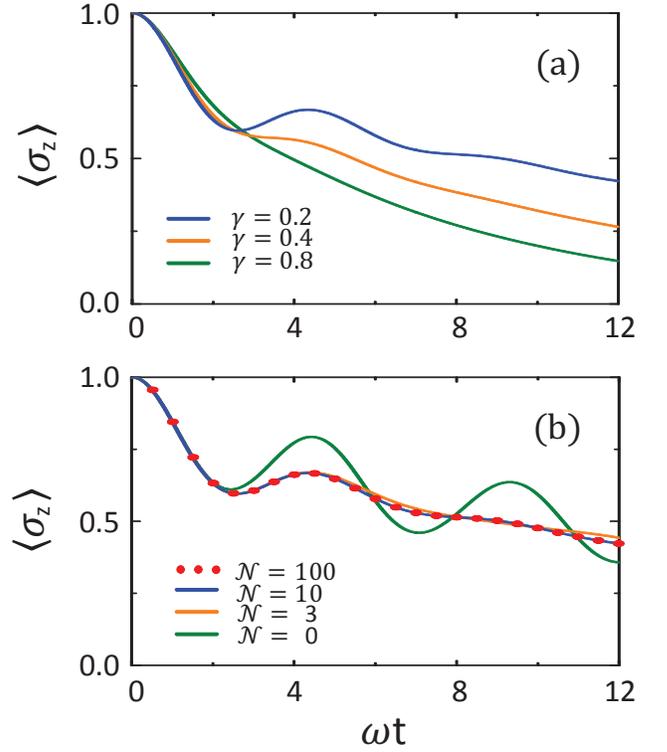}
\caption{(color online)~Time evolution of $\langle\sigma_z\rangle$ for a bath with Ornstein-Uhlenbeck correlation function $\alpha(t,s)=\frac{\Gamma\gamma}{2}e^{-\gamma|t-s|}$.
(a)~$\mathcal{N}=100$, and the inverse of the correlation time is chosen to be $\gamma=0.2$, $0.4$ and $0.8$, respectively.
(b)~$\gamma=0.2$, and the hierarchical order is chosen to be $\mathcal{N}=0$, 3, 10 and 100, respectively.
Also, the coupling strength is chosen to be $\gamma\Gamma=0.2$ in both (a) and (b).}
\label{fig1}
\end{figure}

Many years ago, a proposalwasmade to convert aQLE with
correlated fluctuations into a set of coupled equations \cite{Gri1,Gri2,Gri3}.
It was originally developed to study the QLE with an additive
noise (e.g., the quantum Brownian motion is such a case)
and then extended to the multiplicative-noise case \cite{Gri2}. Here
we study quantum dynamics of the spin-boson model. This
model involves a multiplicative noise in the QLE and is a
more complex, open problem of quantum statistical physics.
The central point of our approach is to reduce the QLE to a
simple differential equation with no noise variables, i.e., the
quantum Bloch equation in Eq. (20). Moreover, we obtain a
set of coupled equations, as in Refs. \cite{Gri1,Gri2,Gri3}, and then use it
to efficiently calculate $\mathcal{O}_0(t)$ in Eq. (20) without invoking any
noise realizations. This is the key reason our approach has a
high numerical efficiency.

\section{Extension to the case of complex correlation function}

When the correlation function is complex, i.e., $\alpha(t,s)\neq\alpha^*(t,z)$, an extra term $\mathcal{W}(t,z)\equiv(0,\langle{w}_{z}(t)|z\rangle\langle{z}|\rangle,\langle{w}_{y}(t)|z\rangle\langle{z}|\rangle)^T$ is added to Eq.~(\ref{SCNQLE3}):
\begin{equation}
\frac{\partial}{\partial t}\mathcal{A}(t,z)
               =\left[-i\mathcal{H}+\mathcal{L}z_t+\mathcal{L}\bar{\mathcal{O}}(t,z)\right]\mathcal{A}(t,z)+\mathcal{W}(t,z).
\label{SCNQLEcc}
\end{equation}
As the simplest approximation, one can apply a Markovian approximation only to the term $\mathcal{W}(t,z)$ in Eq.~(\ref{SCNQLEcc}) by taking $\alpha(t,s)$ in $w_j(t)$ as $\alpha(t,s)=\delta(t-s)$. Then, $\mathcal{W}(t,z)=0$, and both the same c-number stochastic QLE (\ref{SCNQLE3}) and the same non-Markovian quantum Bloch equation (\ref{exact}) are thus obtained.

Also, we can replace $\sigma_x(s)$ in $w_j(t)$ by $\sigma_x(t)$. Then, %we have
$w_z(t) \approx -iv(t)\sigma_y(t)$, and $w_y(t) \approx iv(t)\sigma_z(t)$,
with $v(t)\equiv 4\int_0^tds{\rm Im}\{\alpha(t,s)\}$. This approximation can give very accurate results at the early stage of quantum evolution.
The term $\mathcal{W}(t,z)$ in Eq.~(\ref{SCNQLEcc}) can be written as
\begin{equation}
\mathcal{W}(t,z)=-i\mathcal{V}(t)\mathcal{A}(t,z),
\end{equation}
with
\begin{equation}%{\nonumber}
\mathcal{V}(t)=\left(
\begin{array}{ccc}
0&0&0\\
0&v(t)&0\\
0&0&-v(t)\\
\end{array}\right).
\end{equation}
Thus, Eq.~(\ref{SCNQLEcc}) is reduced to
\begin{equation}
\frac{\partial}{\partial t}\mathcal{A}(t,z)
               =\left[-i\mathcal{H}(t)+\mathcal{L}z_t+\mathcal{L}\bar{\mathcal{O}}(t,z)\right]\mathcal{A}(t,z),
\end{equation}
which has the same form as Eq.~(\ref{SCNQLE3}), with only $\mathcal{H}$ replaced by
$\mathcal{H}(t)=\mathcal{H}+\mathcal{V}(t)$.  Also, we can derive the equation for $\mathcal{O}(t,s,z)$ operator and the non-Markovian quantum Bloch equation, which have the same forms as Eqs.~(\ref{SCNQLE3}) and (\ref{exact}), respectively, but with  $\mathcal{H}$ replaced by $\mathcal{H}(t)=\mathcal{H}+\mathcal{V}(t)$ as well.

\section{Finite-temperature extension}

With the thermo-field method~\cite{Diosi,finitet,finitet2}, we can map the finite-temperature bath onto a larger zero-temperature bath, where a fictitious bath with Hamiltonian $H_{\rm b}=\sum_{k}(-\omega_{k})b_{k}^\dagger b_{k}$ is introduced. The corresponding Hamiltonian of the total system then reads
\begin{equation}
\tilde{H}= \frac{\omega}{2}\sigma_{z}+ \sum_{k}\sigma_{x}\left(g_{k}^*a_{k}^\dagger+ g_{k}a_{k}\right)+ \sum_{k}\omega_{k}\left(a_{k}^\dagger a_{k}- b_{k}^\dagger b_{k}\right).
\end{equation}
When applying a Bogoliubov transformation \cite{finitet2} to the system,
\begin{eqnarray}
a_{k} &=& \sqrt{\bar{n}_{k}+1}c_{k}+ \sqrt{\bar{n}_{k}}d_{k}^\dagger,\nonumber\\
b_{k} &=& \sqrt{\bar{n}_{k}+1}d_{k}+ \sqrt{\bar{n}_{k}}c_{k}^\dagger,
\end{eqnarray}
where $\bar{n}_{k}= [e^{\omega_{k}/k_BT}-1]^{-1}$, the composite bath of bosonic operators $a_k$ and $b_k$ initially prepared in a thermal state is equivalently converted to a virtual composite bath of bosonic operators $c_k$ and $d_k$ in the vacuum state $|0\rangle= |0\rangle_{c}\otimes |0\rangle_{d}$, with $c_{k}|0\rangle_{c}= 0$ and $d_{k}|0\rangle_{d}= 0$. Now, the Hamiltonian of the total system is transformed to
\begin{eqnarray}
\tilde{H} &=& \frac{\omega}{2}\sigma_{z}+ \sum_{k}\sqrt{\bar{n}_{k}+1}\sigma_{x}\left(g_{k}^*c_{k}^\dagger+ g_{k}c_{k}\right)+ \sum_{k}\omega_{k}c_{k}^\dagger c_{k}\nonumber\\
&& +\sum_{k}\sqrt{\bar{n}_{k}}\sigma_{x}\left(g_{k}^*d_{k}+ g_{k}d_{k}^\dagger\right)- \sum_{k}\omega_{k}d_{k}^\dagger d_{k}.
\end{eqnarray}
Similar to Eq.~(\ref{QLE1}), the Pauli operators obeys the QLE
\begin{eqnarray}
\frac{d}{dt}\sigma_x(t) &=& -\omega\sigma_y(t),\nonumber\\
\frac{d}{dt}\sigma_y(t) &=& \omega\sigma_x(t)-2\sigma_z(t)\left[\xi_{T}(t)+\xi_{T}^\dag(t)\right]+w_{Tz}(t),\nonumber\\
\frac{d}{dt}\sigma_z(t) &=& 2\sigma_y(t)\left[\xi_{T}(t)+\xi_{T}^\dag(t)\right]-w_{Ty}(t),
\end{eqnarray}
with the temperature-dependent noise operator
%\begin{equation}
\[
\xi_{T}(t)=\sum_{k}\left[\sqrt{\bar{n}_{k}+1}g_{k}e^{-i\omega_{k}t}c_{k}(0)+ \sqrt{\bar{n}_{k}}g_{k}^*e^{i\omega_{k}t}d_{k}(0)\right],
\]
%\end{equation}
and
%\begin{equation}
\[
w_{Tj}(t)=2i\sigma_j(t)\int_0^tds[\alpha_{T}(t,s)-\alpha_{T}^*(t,s)]\sigma_x(s),
\]
%\end{equation}
where $j=y,z$, and
%\begin{equation}
\[
\alpha_{T}(t,s)= \sum_{k}|g_{k}|^2\left[(\bar{n}_{k}+1)e^{-i\omega_{k}(t-s)}+ \bar{n}_{k}e^{i\omega_{k}(t-s)}\right]
\]
%\end{equation}
is the finite-temperature bath correlation function.

 Using a similar procedures above, we can derive the c-number stochastic QLE at a finite-temperature as
\begin{equation}
\frac{\partial}{\partial t}\mathcal{A}(t,\chi)
               =\left[-i\mathcal{H}+\mathcal{L}\chi_t+\mathcal{L}\bar{\mathcal{O}}(t,\chi)\right]\mathcal{A}(t,\chi)+\mathcal{W}_{T}(t,\chi),
\label{SCNQLET}
\end{equation}
where
%\begin{equation}
\[
\chi_{t}= \sum_{k}\left[\sqrt{\bar{n}_{k}+1}g_{k}e^{-i\omega_{k}t}z_{k}+ \sqrt{\bar{n}_{k}}g_{k}^*e^{i\omega_{k}t}w_{k}\right]
\]
%\end{equation}
is the temperature-dependent noise, and
\[
\bar{\mathcal{O}}(t,\chi)=\int_0^tds\alpha_{T}(t,s)\mathcal{O}(t,s,\chi).
\]
The term $\mathcal{W}_{T}$ in Eq.~(\ref{SCNQLET}) is
\[
\mathcal{W}_{T}(t,\chi)=(0,\langle{w}_{Tz}(t)|zw\rangle\langle{zw}|\rangle,
\langle{w}_{Ty}(t)|zw\rangle\langle{zw}|\rangle)^T,
\]
where $|zw\rangle\equiv|z\rangle\bigotimes|w\rangle$, with the
Bargmann coherent states defined by
\begin{eqnarray}
|z\rangle\equiv \bigotimes_{k}| z_{k}\rangle &=& e^{\sum_{k}z_{k}c_{k}^\dagger}|0\rangle_c, \nonumber\\
|w\rangle\equiv \bigotimes_{k}| w_{k}\rangle &=& e^{\sum_{k}w_{k}d_{k}^\dagger}|0\rangle_d, %\nonumber
\end{eqnarray}
which satisfy $c_k|z\rangle=z_k|z\rangle$ and $d_k|w\rangle=w_k|w\rangle$, respectively.
Note that Eq.~(\ref{SCNQLET}) is formally similar to Eqs.~(\ref{SCNQLE3}) and (\ref{SCNQLEcc}). Therefore, we can solve the finite-temperature problem in an analogous way.

\section{Conclusion}

We have developed a quantum Langevin approach to solving non-Markovian quantum dynamics of the spin-boson model.
Instead of directly attacking the spin-boson model with our non-Markovian QLE, we arrive at a c-number stochastic QLE through projecting the operator QLE onto the coherent states of the bath. Furthermore, we have shown that the stochastic QLE can be reduced to a non-Markovian quantum Bloch equation. With the noiseless quantum Bloch equation, we can efficiently solve the
non-Markovian quantum dynamics of the spin-boson model. In addition, we show that our approach is general enough to include the finite-temperature bath.  Since the spin-boson model
does not admit a non-Markovian master equation, therefore, generally one cannot arrive at a set of useful Bloch equations desirable from our experience in dealing with Markov
systems. We show in this paper that QLE paves a new avenue to bypass the stringent difficulty in deriving the non-Markovian master equations. We expect our stochastic quantum
Langevin approach can play an important role for many other open quantum systems.

%\section*{ACKNOWLEDGMENTS}
\begin{acknowledgments}
This work is supported by the National Natural Science Foundation of China No.~91421102, the National Basic Research Program of China No.~2014CB921401, and the NSAF No.~U1330201. T.Y. is supported by the DOD/AF/AFOSR No.~FA9550-12-1-0001, and he thanks CSRC for hospitality during his visit.
\end{acknowledgments}

%\vspace*{-0.3in}
\begin{widetext}
\appendix
\section{Derivation of the hierarchical equations in Eq.~(\ref{eforQn})}

From the equation of $\mathcal{O}(t,s,z)$ operator in Eq.~(\ref{eforO}), it was obtained~\cite{Yu1} that the $\mathcal{O}_{n}(t,s,v_1,\ldots,v_n)$ operators in Eq.~(\ref{Oexp}) obey the following hierarchical equation:
\begin{eqnarray}
\frac{\partial}{\partial t}\mathcal{O}_{n}(t,s,v_1,\ldots,v_n)
&=& -[i\mathcal{H},\mathcal{O}_{n}(t,s,v_1,\ldots,v_n)]
+(n+1)\mathcal{L}\bar{\mathcal{O}}_{n+1}(t,s,v_1,\ldots,v_n)\nonumber\\
&&+\frac{1}{n!}\sum_{P_n\in S_n}\sum_{k=0}^n[\mathcal{L}\bar{\mathcal{O}}_{k}(t,v_{P_n(1)},\ldots,v_{P_n(k)}),
\mathcal{O}_{n-k}(t,s,v_{P_n(k+1)},\ldots,v_{P_n(n)}],
\label{eforOn}
\end{eqnarray}
with $\mathcal{O}_{0}(t,t)=\mathcal{L}$, $\mathcal{O}_{n}(t,t,v_1,\ldots,v_n)=0$,
and $\mathcal{O}_{n}(t,s,t,v_1,\ldots,v_{n-1})=\frac{1}{n}[\mathcal{L},\mathcal{O}_{n-1}(t,s,v_1,\ldots,v_{n-1})]$ for $n\geq 1$.
Here $S_{n}$ denotes the permutation of all $P_n(k)$'s and $\bar{\mathcal{O}}_{n}(t,v_1,\ldots,v_n)=\int_0^tds\alpha(t,s)\mathcal{O}_{n}(t,s,v_1,\ldots,v_n)$.

Let us define an operator
\begin{equation}
\mathcal{Q}_{n}(t)=\int_0^tds\int_0^tdv_1\ldots \int_0^tdv_n\mathcal{O}_{n}(t,s,v_1,\ldots,v_n)\alpha(t,s)\alpha(t,v_1)\ldots\alpha(t,v_n),
\label{Q}
\end{equation}
with $\mathcal{Q}_0(t)=\bar{\mathcal{O}}_0(t)$, and consider a noise characterized by the Ornstein-Uhlenbeck correlation function $\alpha(t,s)=\frac{\Gamma\gamma}{2}e^{-\gamma|t-s|}$. It can be derived that
\begin{eqnarray}
\frac{\partial}{\partial t}{\bar{\mathcal{O}}}_{0}(t)
&=& \frac{\partial}{\partial t}\int_0^t{ds}\mathcal{O}_{0}(t,s)\alpha(t,s)\nonumber\\
&=& \int_0^tds\left[\frac{\partial}{\partial t}\mathcal{O}_{0}(t,s)\right]\alpha(t,s)+\mathcal{O}_{0}(t,t)\alpha(t,t)
-\gamma\int_0^t{ds}\mathcal{O}_{0}(t,s)\alpha(t,s)\nonumber\\
&=& \int_0^tds\left[\frac{\partial}{\partial t}\mathcal{O}_{0}(t,s)\right]\alpha(t,s)+\frac{\Gamma\gamma}{2}\mathcal{L}
-\gamma\bar{\mathcal{O}}_{0}(t).
\label{Q0}
\end{eqnarray}
From Eq.~(\ref{eforOn}), it follows that $\frac{\partial}{\partial t}\mathcal{O}_{0}(t,s)=-[i\mathcal{H},\mathcal{O}_{0}(t,s)]+\mathcal{L}\bar{\mathcal{O}}_{1}(t,s)
+[\mathcal{L}\bar{\mathcal{O}}_{0}(t),\mathcal{O}_{0}(t,s)]$. Substituting it into Eq.~(\ref{Q0}), we have
\begin{equation}
\frac{\partial}{\partial t}{\bar{\mathcal{O}}}_{0}(t)
=-[i\mathcal{H},\bar{\mathcal{O}}_{0}(t)]+[\mathcal{L}\bar{\mathcal{O}}_{0}(t),\bar{\mathcal{O}}_{0}(t)]
-\gamma\bar{\mathcal{O}}_{0}(t)+\frac{\Gamma\gamma}{2}\mathcal{L}+\mathcal{L}\mathcal{Q}_{1}(t).
\label{eforQ0}
\end{equation}
This is the first equation in Eq.~(\ref{eforQn}).

For $n\geq1$, it can be derived that
\begin{eqnarray}
\frac{\partial}{\partial t}\mathcal{Q}_{n} &=& \frac{\partial}{\partial t}\int_0^tds\int_0^tdv_1\ldots \int_0^tdv_n\mathcal{O}_{n}(t,s,v_1,\ldots,v_n)\alpha(t,s)\alpha(t,v_1)\ldots\alpha(t,v_n)\nonumber\\
&=&\int_0^tds\int_0^tdv_1\ldots \int_0^tdv_n\left[\frac{\partial}{\partial t}\mathcal{O}_{n}(t,s,v_1,\ldots,v_n)\right]\alpha(t,s)\alpha(t,v_1)\ldots\alpha(t,v_n)\nonumber\\
&&+\alpha(t,t)\int_0^tdv_1\ldots \int_0^tdv_{n-1}\mathcal{O}_{n}(t,t,v_1,\ldots,v_{n-1})\alpha(t,v_1)\ldots\alpha(t,v_{n-1})\nonumber\\
&&+n\alpha(t,t)\int_0^tds\int_0^tdv_1\ldots \int_0^tdv_{n-1}\mathcal{O}_{n}(t,s,t,v_1,\ldots,v_{n-1})\alpha(t,s)\alpha(t,v_1)\ldots\alpha(t,v_{n-1})\nonumber\\
&&+\int_0^tds\int_0^tdv_1\ldots \int_0^tdv_n\mathcal{O}_{n}(t,s,v_1,\ldots,v_n)\left[\frac{\partial}{\partial t}\alpha(t,s)\alpha(t,v_1)\ldots\alpha(t,v_n)\right] \nonumber\\
&=& \int_0^tds\int_0^tdv_1\ldots \int_0^tdv_n\left[\frac{\partial}{\partial t}\mathcal{O}_{n}(t,s,v_1,\ldots,v_n)\right]\alpha(t,s)\alpha(t,v_1)\ldots\alpha(t,v_n)\nonumber\\
&&+\frac{\Gamma\gamma}{2}[\mathcal{L},\mathcal{Q}_{n-1}]-(n+1)\gamma \mathcal{Q}_{n},
\label{Qn}
\end{eqnarray}
where we have used the relations $\mathcal{O}_{n}(t,t,v_1,\ldots,v_n)=0$,
and $\mathcal{O}_{n}(t,s,t,v_1,\ldots,v_{n-1})=\frac{1}{n}[\mathcal{L},\mathcal{O}_{n-1}(t,s,v_1,\ldots,v_{n-1})]$.
Substituting $\frac{\partial}{\partial t}\mathcal{O}_{n}(t,s,v_1,\ldots,v_n)$ in Eq.~(\ref{eforOn}) into Eq.~(\ref{Qn}), we then obtain
\begin{equation}
\frac{\partial}{\partial t}\mathcal{Q}_{n}=-i[\mathcal{H},\mathcal{Q}_{n}]+\sum_{k=0}^n[\mathcal{L}\mathcal{Q}_{k},\mathcal{Q}_{n-k}]
-(n+1)\gamma \mathcal{Q}_{n}+\frac{\Gamma\gamma}{2}[\mathcal{L},\mathcal{Q}_{n-1}]
+(n+1)\mathcal{L}\mathcal{Q}_{n+1},
\end{equation}
which is the second equation in Eq.~(\ref{eforQn}).

\end{widetext}

\end{document}